          [2001/04/25 1.1 (PWD)]
\documentclass[a4paper,twocolumn]{esapub} 

\usepackage{times}
\usepackage{amsmath,amssymb}
\usepackage{epsfig}

\title{X-RAY CHEMISTRY IN THE ENVELOPES AROUND YOUNG STELLAR OBJECTS}
\author{Pascal St\"auber}
\author{A. O. Benz}
\affil{\textit{Institute of Astronomy, ETH-Zentrum, CH-8092 Zurich, Switzerland,
Email: pascalst@astro.phys.ethz.ch, benz@astro.phys.ethz.ch}}
\author{S. D. Doty}
\affil{\textit{Department of Physics and Astronomy, Denison University, Granville,
 OH 43023, USA, Email: doty@cc.denison.edu}}
\author{E. F. van Dishoeck}
\affil{\textit{Sterrewacht Leiden, PO Box 9513, 2300 RA Leiden, The Netherlands, Email:
ewine@strw.leidenuniv.nl}}

\begin{document}


\maketitle
\thispagestyle{empty}

\begin{abstract}
We have studied the influence of X-rays from a massive young stellar object 
(YSO) on the chemistry of its own envelope by extending the models of Doty 
et al. (2002) and St\"auber et al. (2004a). The models are applied to the 
massive star-forming region AFGL 2591 for different X-ray luminosities and 
plasma temperatures. Enhanced column densities for several species are 
predicted. In addition we present first detections of CO$^+$ and SO$^+$ toward 
AFGL 2591.These molecular ions are believed to be high-energy tracers. 
Herschel-HIFI will be able to observe other tracers like CH and CH$^+$ 
whereas ALMA is well suited to measure the size and geometry of the emitting
region. 
\end{abstract}

\section{Introduction}

Observational studies of star-forming regions show that young stellar objects 
(YSOs) are very strong X-ray emitters. Typical X-ray luminosities range from 
approximately 10$^{28}$ to 10$^{33}$\,erg\,s$^{-1}$ with temperatures between 
10$^6$--10$^8$\,K (e.g., Feigelson \& Montmerle 1996; Beuther et al. 2002). In 
the earliest stage of evolution, the protostar is still deeply embedded in its 
natal molecular cloud (A$_V >$ 100\,mag). As a consequence, X-rays are not 
directly observable toward very young objects and the onset of the high energy 
radiation remains a secret to this day. Molecular gas exposed to X-rays forms 
an X-ray dissociation region (XDR) with a peculiar chemistry and physical 
structure (e.g., Krolik \& Kallman 1983; Maloney et al. 1996). Nearly all XDR 
models, however, concentrate on the impact of X-rays on molecular clouds
or on the surroundings of active galactic nuclei (AGNs). To study the influence 
of X-rays on the chemistry in dense YSO envelopes and find molecular tracers for 
high energetic photons, we have extended the chemical models of Doty et al. 
(2002) and St\"auber et al. (2004a). To compare our model calculations to 
observations we have carried out a single-dish survey of molecular ions 
(CO$^+$, SO$^+$) and radicals (CN, NO) toward both low-mass and high-mass 
star-forming regions. The challenge is, however, to distinguish between UV 
tracers (see St\"auber et al. 2004a) and X-ray tracers, since both kinds of 
high-energy radiation tend to form ions and radicals.

\section{Chemical X-ray Model}

The chemical model is based on the detailed thermal and gas-phase chemistry 
models of Doty et al. (2002), including UV radiation from the in- or outside 
(St\"auber et al. 2004a). The model assumes a given temperature and density 
distribution and calculates the time-dependent chemistry at a certain distance 
from the central source. For the chemical X-ray network we follow Maloney et al. 
(1996) and Yan (1997). It contains direct X-ray ionization and dissociation of 
atomic and molecular species and X-ray induced reactions caused by electron 
impact. The main reactions of the electrons are the ionization and dissociation 
of H$_2$ and the excitation of H, He and H$_2$. The electronically excited 
states of H, He and H$_2$ decay back to the ground states by emitting UV 
photons. The internally generated ultraviolet photons can photoionize and 
photodissociate other species in the gas. These secondary processes are far 
more important for the chemical network than the primary interaction of the 
X-rays with the gas, making it more difficult to distinguish X-ray and
UV-induced processes.

We have applied our model to the massive star-forming region AFGL 2591 and
adopt the temperature and power-law density distribution proposed by van der Tak
et al. (1999) and Doty et al. (2002). The most important parameter in the 
chemistry and physics of XDRs is $H_{{\rm X}}/n$, the local X-ray energy 
deposition rate per particle (Maloney et al. 1996). Since the heating rate is 
$\sim$\,proportional to this ratio, we can estimate its importance for the gas 
temperature. Our first radial point of interest is at $\sim$\,200\,AU from the 
central source where the density is already fairly high ($n \sim 10^7$) but 
$H_{{\rm X}}$ is low due to absorption and geometric dilution. We can therefore neglect 
additional heating of the gas through X-rays and assume $T_{{\rm dust}} \approx 
T_{{\rm gas}}$. For the X-ray spectrum we have fitted a thermal spectrum $\propto$ 
exp(-$E/kT_{{\rm X}}$) which is normally used to fit observed spectra toward 
star-forming regions. 

\begin{figure}
\centering
\includegraphics[width=\linewidth]{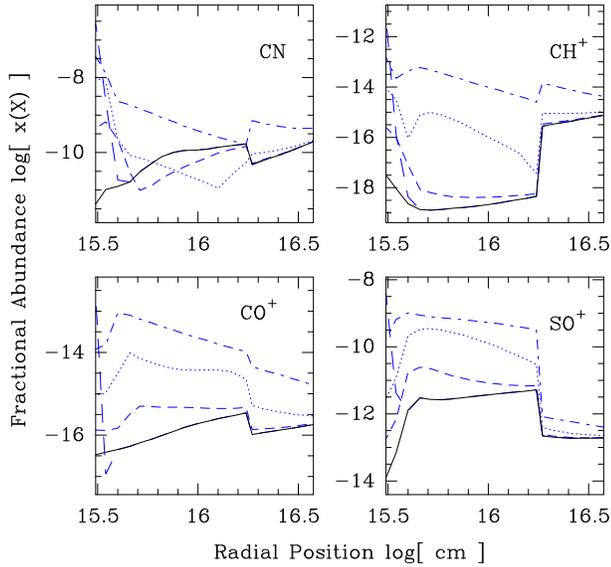}
\caption{Depth dependent fractional abundances of CN, CH$^+$, CO$^+$
and SO$^+$. The solid line corresponds to the model without any high energy
radiation. The long-dashed line is the model without X-rays but with an inner
FUV field, the dashed line represents $L_{{\rm X}} = 10^{30}$\,erg\,s$^{-1}$, the dotted
line $L_{{\rm X}} = 10^{31}$\,erg\,s$^{-1}$ and the dash-dotted line 
$L_{{\rm X}} = 10^{32}$\,erg\,s$^{-1}$.
\label{fig:model}}
\end{figure}

Fig. \ref{fig:model} shows some model results for AFGL 2591 with different X-ray
luminosities for a plasma temperature of 10$^7$\,K. Similar to the 
results of St\"auber et al. (2004a) where the impact of an inner FUV field has
been studied, the abundances of many ions and radicals are enhanced by several 
orders of magnitude. The main difference between the FUV and X-ray models is that 
X-rays can penetrate deeper into the envelope and affect the gas-phase chemistry
even at large distances from the source. Whereas the FUV enhanced species cover a region 
of $\sim$\,200--300\,AU into the cloud, the enhanced region through X-rays is 
$\sim$\,1100\,AU in our models. The reason for this are the small 
cross-sections for X-rays, which become smaller the harder the X-ray photons. 
Species that are enhanced by X-rays are among others CH, CN, HCN, CH$_2$CN, SO, 
CH$^+$, CO$^+$ and SO$^+$.

\section{Observations}
\thispagestyle{empty}

\begin{figure}
\centering
\includegraphics[width=\linewidth]{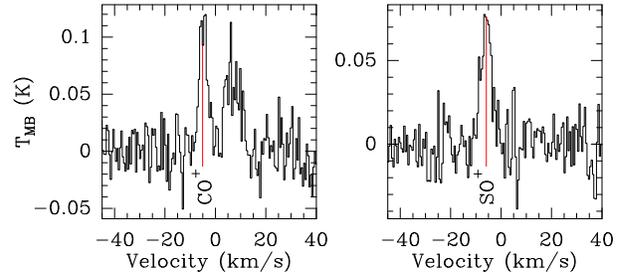}
\caption{The 3--2 transition of CO$^+$ at 353.741\,GHz and SO$^+$
at 347.740\,GHz observed toward AFGL 2591 in 2004 with the JCMT.
\label{fig:lines}}
\end{figure}

In addition to our theoretical models we are also searching for observational 
X-ray and FUV tracers near deeply embedded objects. We have carried out a 
survey of the rotational 3--2 transitions of CN, NO, CO$^+$ and SO$^+$ toward a 
sample of 11 sources -- both high- and low-mass -- with the James Clerk Maxwell 
Telescope on Mauna Kea, Hawaii. These high-J transitions probe uniquely the
dense gas close to the protostar and filter out the lower density material. CN
and NO are detected in all observed sources, but CO$^+$ and SO$^+$ only toward
the high-mass YSOs (see also St\"auber et al. 2004b). CO$^+$ is tentatively 
seen toward one low-mass source, IRAS 16293. Fig. \ref{fig:lines} shows the 
first detections of CO$^+$ and SO$^+$ toward AFGL 2591. 

\section{Future Observations}
\thispagestyle{empty}

Our models have shown that hydrides like CH or CH$^+$ are enhanced due to 
either inner FUV fields (St\"auber et al. 2004a) or due to X-ray emission from 
the central source (St\"auber et al., in prep.). However, these lines are not 
observable with ground based telescopes and are therefore candidates for 
Herschel-HIFI observations. In order to constrain the size and geometry of 
the emitting region of ions like CO$^+$ or SO$^+$ the high angular 
resolution and sensitivity of ALMA are needed. It will then be possible to make 
precise estimates of the X-ray and FUV flux and maps will reveal whether the 
emission is really concentrated in the surrounding envelope or whether it is in 
the outflow cones or even in the protostellar disk. For example, the 0.1\,arcsec 
beam of ALMA represents a resolution of $\sim$\,200\,AU toward AFGL 2591 which 
should be sufficient to resolve the different regions. Such observations will 
provide new and important information about the evolution of the surroundings 
of young stellar objects.


\thispagestyle{empty}
\end{document}